# Heuristic Optimization of Amplifier Reconfiguration Process for Autonomous Driving Optical Networks

Qizhi Qiu, Xiaomin Liu, Yihao Zhang, Lilin Yi, Weisheng Hu, and Qunbi Zhuge*

State Key Laboratory of Advanced Optical Communication Systems and Networks, Department of Electronic Engineering, Shanghai Jiao Tong University, Shanghai, China, *qunbi.zhuge@sjtu.edu.cn

**Abstract** *We propose a heuristic-based optimization scheme for reliable optical amplifier reconfiguration process in ADON. In the experiment on a commercial testbed, the scheme prevents a 1.0-dB Q-factor degradation and outperforms 98.5% random solutions.* ©2024 The Author(s)

**Introduction**

To satisfy the increasing demands of network capacity and reduce the operating expenditure (OPEX), the concept of autonomous driving optical networks (ADON) is introduced to facilitate intelligent management without human intervention [1]. One of ADON's key capabilities is the dynamic optimization for better quality of transmission (QoT). Among multiple optimization tasks, autonomous optical power optimization is particularly essential and has been widely investigated [2-6]. In [3], the signal power of each optical multiplex section (OMS) is dynamically optimized by reconfiguring the booster amplifier. In [4], the optical power along the link is optimized by adjusting the optical amplifier (OA) settings in a wavelength reconfiguration scenario. Moreover, the optimization scenario is extended to C+L band by considering the influence of stimulated Raman scattering [5, 6].

However, previous studies do not pay attention to the influence of the OA reconfiguration process. Generally, due to the controller-to-device latency and distinct response time, it takes different time for the reconfigurations of different OAs. As a result, in real systems, the OAs are usually reconfigured one by one. During this process, the system performance might fluctuate, and undesired disruption to existing services might occur. Therefore, besides optimal OA configuration, it is also necessary to figure out the optimal order so as to avoid the performance fluctuation. However, in real systems with plenty of OAs and multiple configurable OA parameters, determining the optimal reconfiguration order becomes an NP-hard problem.

In this paper, we propose a heuristic-based optimization scheme to achieve a reliable OA reconfiguration process in ADON. First, we construct a digital twin (DT) based on an artificial neural network (ANN) and telemetry data [7, 8]. The DT assists to estimate the QoT variations during the OA reconfiguration process. Then, a heuristic algorithm is utilized to search the optimal OA reconfiguration order. In the experiment on a commercial testbed, the proposed scheme prevents a Q-factor degradation of up to 1.0 dB during the OA reconfiguration process. In addition, the minimum Q-factor during the OA reconfiguration outperforms 98.5% of the random solutions.

**Principle**

Fig. 1(a) illustrates the operation framework of an ADON with the proposed OA reconfiguration order optimizer. When new services need to be established, the software-defined network (SDN) controller manages the service provisioning and then loads the new wavelengths. As the loading scenario changes, the optimal OA configurations change as well, so that optical power optimization of each span should be reconducted to update OA settings. Afterwards, the proposed scheme is employed to determine the optimal reconfiguration order. Specifically, the reconfiguration process includes adjusting the gain and tilt of each OA. Note that, the gain and tilt of each OA can be adjusted separately. Therefore, with $N$ OAs to be reconfigured, the reconfiguration consists of $2N$ steps. Each reconfiguration order is represented as a sequence of numbers from 1 to $2N$, where each number corresponds to a specific reconfiguration step. All the reconfiguration orders will lead to the same final performance. However, during the reconfiguration process when only some of the OAs have been configured, the performance might significantly degrade as shown in our experiments. Therefore, it is essential to search for the optimal reconfiguration order to avoid performance degradation.

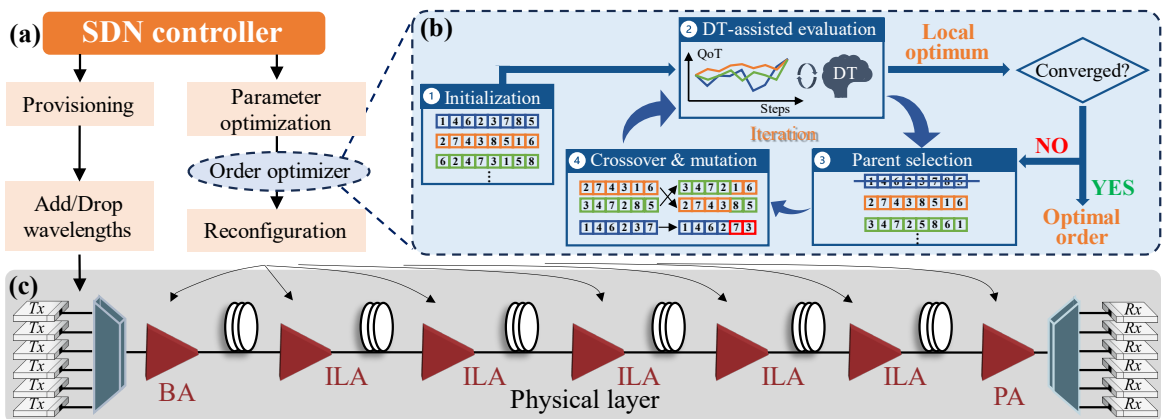

**Fig. 1:** (a) The operation framework of SDN-based ADON. (b) The diagram of the heuristic-based OA reconfiguration order optimizer. (c) Experiment setup.

The diagram of the proposed OA reconfiguration order optimizer is depicted in Fig. 1(b). The scheme is based on the genetic algorithm (GA), including four main steps as follows: (1) *Initialization*. First, different OA reconfiguration orders are randomly generated as the initial population. (2) *DT-assisted evaluation*. Then the performance of the OA reconfiguration orders is evaluated with a DT model. Specifically, the DT is constructed based on a simple ANN and telemetry data including different OA configurations and corresponding Q-factors. In this way, the DT can emulate the reconfiguration processes and record the Q-factor variations. Fitness values are then calculated as the sum of mean and minimum Q-factors. (3) *Parent selection*. Next, the individuals with lower fitness values are eliminated, while the others are selected as parent orders. (4) *Crossover & mutation*. Considering the specific part in the reconfiguration order (e.g., reconfigure the 1st, 2nd, and 3rd OAs' gains successively) may lead to better performance, partially matched crossover (PMX) is applied in the iteration. Note that, PMX may cause repeated configuration steps in an order. Therefore, subsequent corrections will be applied. Additionally, a mutation operation is applied to avoid the premature convergence. The iteration of evaluation, parent selection, PMX and mutation continues until the result converges. Finally, the converged result is output as the optimal order.

**Experiment setup**

The experiment setup of the commercial testbed is depicted in Fig. 1(c). Six transponders operating at 400 Gbps are applied to generate 63.9-Gbaud 16QAM signals. The channel spacing is 75 GHz. The transmission link is composed of six 80-km G.652 fibers, one boost amplifier, five in-line amplifiers, and one pre-amplifier. All the transponders and OAs are managed and controlled by an SDN controller with NETCONF protocol and YANG models. In addition, the gains and tilts of all seven OAs are adjustable through the SDN controller with a step size of 0.1 dB.

As shown in Fig. 2(a), two cases are considered to evaluate the proposed scheme. For each case, two scenarios are emulated in the experiment. The initial loading scenario represents the loading status before new wavelengths are loaded, while the current loading scenario represents the loading status with new wavelengths. For case 1 and case 2, the initial loading scenario contains one and two batches of signals, respectively. Particularly, each batch of signals contains 6 dummy signals and one real signal. The dummy signals are generated from an amplified spontaneous emission (ASE) source and a programmable spectrum processor, which are coupled with real signals after a multiplexer. In the current loading scenario, 6 batches of signals are loaded. In this way, the QoT of each batch can be represented by the Q-factor of the central signal. The configurations with optimal performance in the initial and current loading scenarios are selected from 500 randomly generated sets, as the initial and target configurations, respectively.

The DT is based on an ANN with a single hidden layer. The input feature is normalized gains and tilts of the 7 OAs along the link, while the output comprises Q-factors of the 6 signal batches. 1000 pairs of OA configurations and corresponding Q-factors are collected. 700 pairs are employed for training and the other 300 pairs are used for validation. The mean square error (MSE) loss function is utilized, and the optimizer is set as Adam.

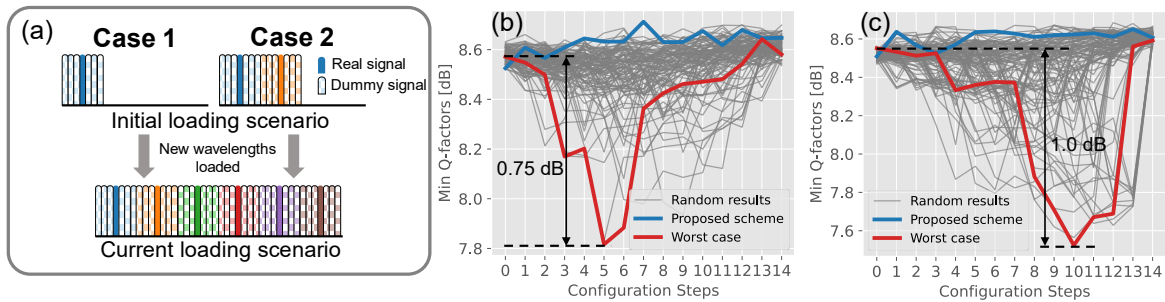

**Fig. 2: (a)** Schematic diagram of case 1 and case 2. **(b)** Q-factor variations of different OA reconfiguration orders for case 1. **(c)** Q-factor variations of different OA reconfiguration orders for case 2.

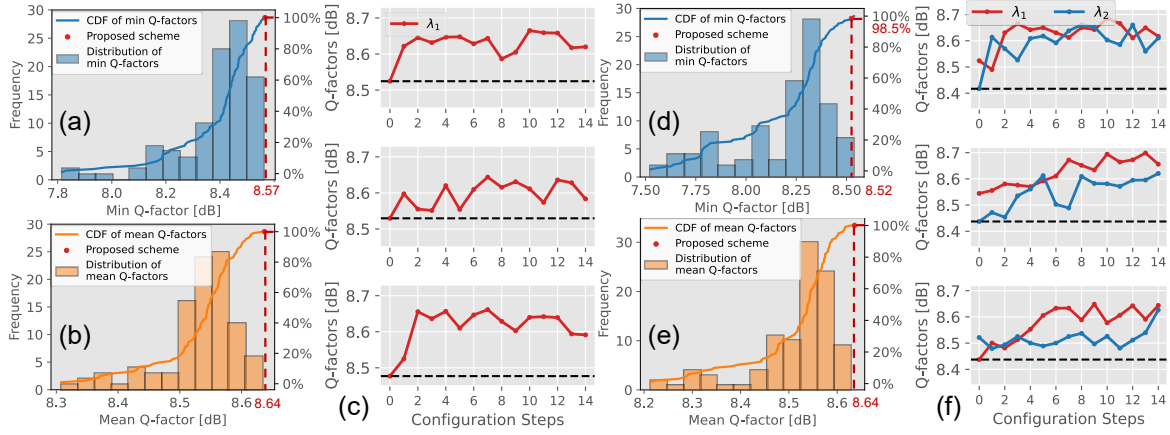

**Fig. 3:** Distribution and CDF of the minimum and mean Q-factors for **(a) (b)** case 1 and **(d) (e)** case 2, and Q-factor variations of the existing signal batches with 3 different initial configurations for **(c)** case 1 and **(f)** case 2, exploiting the proposed scheme.

As for the heuristic algorithm parameters, the population includes 64 random orders from 1 to 14, corresponding to configuration steps of the 7 OAs' gains and tilts. The probability of crossover and mutation is 0.8 and 0.1, respectively. In addition, the convergence condition is that the local optimum remains the same for 20 generations.

**Results and discussions**
For comparison, 100 sets of random orders are generated for each case. Specifically, the OAs in the testbed are firstly set to the initial configuration, and then reconfigured to the target configuration in orders generated randomly. The Q-factor variations during the process are recorded for evaluation. In Fig. 2(b), the Q-factors of the existing signal batch in case 1 are depicted for each configuration step. Particularly, the results of the proposed scheme and the worst case are respectively highlighted in blue and red. During the OA reconfigurations, Q-factors of the proposed scheme is always higher than that at step 0, indicating the prevention of Q-factor degradation, which is up to 0.75 dB in the experiment. Similarly in Fig. 2(c), the minimum Q-factors of the existing two signal batches in case 2 are shown. The proposed scheme also effectively prevents the Q-factor degradation, which is up to 1.0 dB in the experiment. It is worth noting that, the deviations of Q-factors at step 0 and step 14 are caused by the performance fluctuation of the experimental testbed under the same OA configuration, typically within 0.1 dB.

Moreover, the minimum and mean Q-factor distributions of the 100 random orders for case 1 are shown as histograms in Fig. 3(a) and 3(b), and the cumulative distribution function (CDF) curve is presented as well. The result of the proposed scheme is marked on the CDF curve, which outperforms 100% of the random instances for case 1. Additionally, the proposed scheme is evaluated in three more scenarios, each with distinct initial configurations but the same target configuration. Utilizing the proposed scheme, Q-factor degradations during the OA reconfiguration are effectively prevented in these scenarios, as shown in Fig. 3(c).

For case 2, the proposed scheme achieves a minimum Q-factor better than 98.5% of the results, with the mean Q-factor outperforming all the instances, as shown in Fig. 3(d) and 3(e). The scheme is also applied in three scenarios with different initial configurations but the same target configuration. The Q-factor variations in these scenarios are depicted in Fig. 3(f), demonstrating that the Q-factor degradation is effectively prevented with the proposed order optimization scheme.

**Conclusion**
In this paper, we propose a heuristic-based order optimization scheme for reliable OA reconfiguration in ADON. Leveraging a heuristic algorithm and a DT, the proposed scheme efficiently generates near-optimal OA reconfiguration orders from a huge solution space. In the experiment on a commercial testbed, the proposed scheme prevents a Q-factor degradation of up to 1.0 dB. The results also demonstrate its superiority over 98.5% of randomly generated solutions.

**Acknowledgement**
This work was supported by Shanghai Pilot Program for Basic Research-Shanghai Jiao Tong University (21TQ1400213) and National Natural Science Foundation of China (62175145).